%% file: main.tex
\documentclass{article}
\usepackage{spconf,amsmath,graphicx}
\usepackage{algorithm, algorithmic}
\usepackage{amssymb}
\usepackage[colorlinks, linkcolor=black, anchorcolor=black, citecolor=black]{hyperref}

\usepackage{textcomp}
\usepackage{amsthm}
\usepackage{xcolor}
\usepackage{units}
\usepackage{balance}
\usepackage{acro}
\include{Acronyms}

\usepackage{bm}
\usepackage{tikz} 
\usepackage[utf8]{inputenc}
\usepackage{pgfplots} 
\usepackage{pgfgantt}
\usepackage{pdflscape}
\usepackage{hyperref}
\hypersetup{urlcolor=black}

\DeclareMathOperator*{\argmax}{arg\,max}
\DeclareMathOperator*{\argmin}{arg\,min}

\title{Misspecified Cram\'er-Rao Bound of RIS-aided Localization under Geometry Mismatch\\
\thanks{T\lowercase{his work is supported by \uppercase{KAUST} \uppercase{O}ffice of \uppercase{S}ponsored \uppercase{R}esearch (\uppercase{OSR}) under \uppercase{A}ward \uppercase{N}o. \uppercase{ORA-CRG2021-4695}, 
and by the \uppercase{E}uropean \uppercase{C}ommission through the \uppercase{EU H2020 RISE-6G} project under grant 101017011.}
}
}
\name{Pinjun Zheng$^\star$, 
Hui Chen$^\dagger$,
Tarig Ballal$^\star$, 
Henk Wymeersch$^\dagger$, 
Tareq Y. Al-Naffouri$^\star$}
\address{$^\star$King Abdullah University of Science and Technology, Thuwal, KSA\\
$^\dagger$Chalmers University of Technology, Gothenburg, Sweden}

\begin{document}
%
\maketitle
\begin{abstract}
In 5G/6G wireless systems, reconfigurable intelligent surfaces (RIS) can play a role as a passive anchor to enable and enhance localization in various scenarios. 
However, most existing RIS-aided localization works assume that the geometry of the RIS is perfectly known, which is not realistic in practice due to calibration errors. 
In this work, we derive the misspecified Cram\'er-Rao bound (MCRB) for a single-input-single-output RIS-aided localization system with RIS geometry mismatch. 
Specifically, unlike most existing works that use numerical methods, we propose a closed-form solution to the pseudo-true parameter determination problem for MCRB analysis.
Simulation results demonstrate the validity of the derived pseudo-true parameters and MCRB,
and show that the RIS geometry mismatch causes performance saturation in the high signal-to-noise ratio regions.

\end{abstract}
\section{Introduction}

\Ac{ris}-aided localization has been extensively studied recently~\cite{wymeersch2020radio,zhang2022toward,Alghamdi2020Intelligent}. One of the merits is that \ac{ris} enables localization in extreme scenarios. For example, a \ac{siso} system consisting of a \ac{bs} and a \ac{ue} can perform communication, but achieving localization is impossible. With the introduction of the RIS channel and the corresponding delay and \ac{aod} measurements, however, joint \ac{ue} localization and synchronization can be completed~\cite{keykhosravi2021siso}. RIS is also beneficial in other scenarios such as wireless fingerprinting localization~\cite{nguyen2021wireless}, signal strength-based localization~\cite{zhang2021metaLocalization}, localization under mobility~\cite{keykhosravi2022ris}, user tracking~\cite{teng2022bayesian}, terahertz band localization~\cite{Sarieddeen2021Overview}, etc. 

Unfortunately, almost all the mentioned systems assume that the \ac{ris} position and orientation are perfectly known. However, the geometry mismatch of the \ac{ris} is likely to be introduced in reality due to calibration errors. These calibration errors propagate to the estimated channel parameters and cause the model mismatch in the position estimation process. 
For the mismatched model, the \ac{mcrb} is a tool to quantify the impact of the \ac{ris} calibration error~\cite{fortunati2017performance}.
When using \ac{mcrb}, the assumed channel model is different from the true model, and a misspecified performance bound can be derived with the model mismatch considered.
The \ac{mcrb} analysis for radio localization under hardware impairment~\cite{ozturk2022impact} and channel model mismatch~\cite{chen2022channel} have been reported in previous works. In the \ac{mcrb} derivation, one of the most essential steps is determining the pseudo-true parameters~\cite{fortunati2017performance}, which is to find a solution that minimizes the \ac{kld} between the true and mismatched statistical models and is usually accomplished using numerical methods~\cite{ozturk2022impact,chen2022channel}. 
In this work, we aim to derive the \ac{mcrb} to evaluate the impact of \ac{ris} geometry mismatch on RIS-assisted localization.
The main contribution of this work is that we derive a \emph{closed-form} solution to the pseudo-true parameter determination problem. The corresponding geometrical interpretation is also given.
The simulation code of this paper is available at  \url{https://github.com/ZPinjun/RISgeoMCRB2023ICASSP}.

\section{System Model}
\subsection{Geometrical Relations} 
\begin{figure}[t]
    \centering
    \includegraphics[width=2.6in]{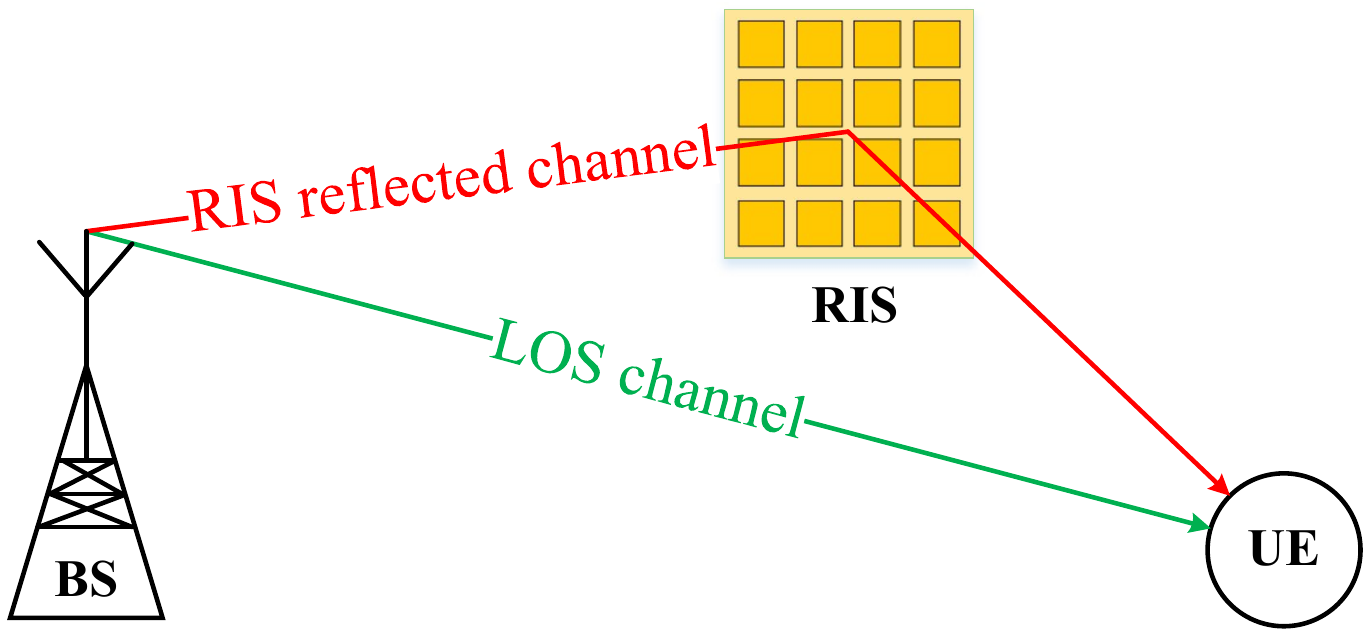}
    \vspace{-1em}
    \caption{ 
        Illustration of a \ac{ris}-aided localization system. 
      }
    \label{fig_system}
    \vspace{-4mm}
\end{figure}

We consider a downlink \ac{siso} wireless system with a \ac{bs}, a \ac{ue}, and a \ac{ris}, as shown in Fig.~\ref{fig_system}.
We indicate the position of the \ac{bs} and the \ac{ue} by $\mathbf{p}_\text{b}\in\mathbb{R}^3$ and $\mathbf{p}\in\mathbb{R}^3$.
The position and orientation of the \ac{ris} are denoted as $\mathbf{p}_\text{r}\in\mathbb{R}^3$ and $\mathbf{R}_\text{r}\in \text{SO(3)}$~\cite{absil2009optimization}, respectively.
The entries of $\mathbf{p}_\text{b}$, $\mathbf{p}_\text{r}$, and $\mathbf{R}_\text{r}$ are assumed to be known.
Besides, we assume an unknown clock bias $\Delta\in\mathbb{R}$ exists between the \ac{ue} and the \ac{bs}~\cite{Bjornson2021Reconfigurable,zheng2022coverage}.

In the \ac{ris}'s \ac{lcs}, the \ac{aoa} from the \ac{bs} consists of an azimuth angle $\theta_{\text{az}}$ and an elevation angle $\theta_{\text{el}}$, 
while the \ac{aod} towards the \ac{ue} consists of an azimuth angle $\phi_{\text{az}}$  and an elevation angle $\phi_{\text{el}}$.
For compactness, we define $\bm{\theta}=[\theta_{\text{az}},\theta_{\text{el}}]^\mathsf{T} $ and $\bm{\phi}=[\phi_{\text{az}},\phi_{\text{el}}]^\mathsf{T} $.
Since we assume the geometry of the \ac{bs} and the \ac{ris} to be known in the ideal model, we focus on the \ac{aod} towards the \ac{ue} with an unknown position, which is given by
\begin{align}
\label{eq_phi}
  \phi_\text{az} &= \arctan 2\left([\mathbf{R}_\text{r}^\mathsf{T} (\mathbf{p}-\mathbf{p}_\text{r})]_2, [\mathbf{R}_\text{r}^\mathsf{T} (\mathbf{p}-\mathbf{p}_\text{r})]_1\right),
  \\
  \phi_\text{el} &= \arcsin\left({[\mathbf{R}_\text{r}^\mathsf{T} (\mathbf{p}-\mathbf{p}_\text{r})]_3}/{\|\mathbf{p}-\mathbf{p}_\text{r}\|}\right),
\end{align}
where $[\cdot]_i$ represents the $i$-th element of a vector.
The delays over the \ac{bs}-\ac{ue} \ac{los} and the \ac{bs}-\ac{ris}-\ac{ue} \ac{nlos} paths are given by
\begin{align}
  \tau_\text{b} &= {\|\mathbf{p}_\text{b}-\mathbf{p}\|}/{c} + \Delta, \\
  \tau_\text{r} &= {\|\mathbf{p}_\text{b}-\mathbf{p}_\text{r}\|/{c} + \|\mathbf{p}_\text{r}-\mathbf{p}\|}/{c} + \Delta,
  \label{eq_tau}
\end{align}
where $c$ is the speed of light.


\subsection{Channel Model}
We consider the transmission of $L$ \ac{ofdm} pilot symbols with $K$ subcarriers.
The frequency of the $k$-th subcarrier is denoted as $f_k= f_c + \frac{(2k-1-K)}{2}\Delta_f, k=1,\dots,K$, where $\Delta_f=B/K$ is the subcarrier spacing, $f_c$ is the carrier frequency, and $B$ is the bandwidth.
The received baseband signal in the \ac{ofdm} block with index $\ell$ is given by~\cite{Bjornson2021Reconfigurable} 
\vspace{-1em}
\begin{equation}\label{eq_channel}
  \mathbf{y}_\ell = \underbrace{ \overbrace{g_\text{b}\mathbf{d}(\tau_\text{b})\odot\mathbf{x}}^{\text{\ac{los} channel}} + \overbrace{g_\text{r}\mathbf{b}(\bm{\phi})^\mathsf{T}\bm{\gamma}_\ell(\mathbf{d}(\tau_\text{r})\odot\mathbf{x})}^{\text{\ac{ris} reflected channel}}}_{\bm{\mu}_\ell} + \mathbf{n},
  \vspace{-1em}
\end{equation}
where $g_\text{b}$ and $g_\text{r}$ indicate the complex channel gain for the \ac{los} path and the \ac{ris} reflected path, $\bm{\gamma}_\ell$ is the \ac{ris} phase profile, $\mathbf{x}$ is the transmitted signal,
$\mathbf{n}\sim\mathcal{CN}(\mathbf{0},\sigma^2\mathbf{I}_K)$ is a complex \ac{awgn} vector, and $\bm{\mu}_\ell$ is the noise-free version of the received signal.
The vector $\mathbf{d}(\tau)$ is the delay steering vector defined as
\begin{equation}
  \mathbf{d}(\tau) = [1, e^{-j2\pi\Delta_f\tau}, \dots, e^{-j2\pi(K-1)\Delta_f\tau}]^\mathsf{T}.
\end{equation}
Moreover, $\mathbf{b}(\bm{\phi})\triangleq\mathbf{a}(\bm{\theta})\odot\mathbf{a}(\bm{\phi})$,
where $\mathbf{a}(\bm{\theta})$ and $\mathbf{a}(\bm{\phi})$ are respectively the array response vectors for the \ac{aoa} and \ac{aod} of the \ac{ris}.
Under the narrowband far-field model, the array response vector of the \ac{ris} can be described as
$
  \left[\mathbf{a}\left(\bm{\alpha}\right)\right]_i = e^{j\frac{2\pi f_c}{c}\mathbf{t}(\bm{\alpha})^\mathsf{T}\mathbf{p}_i},
$
where $\mathbf{p}_i$ is the position of the $i$-th \ac{ris} element given in its \ac{lcs}, and $\mathbf{t}(\bm{\alpha})$ is the direction vector defined as 
\begin{equation}\notag
  \mathbf{t}(\bm{\alpha}) \triangleq 
\begin{bmatrix}
  \cos(\alpha_{\text{az}})\cos(\alpha_{\text{el}}),\ 
  \sin(\alpha_{\text{az}})\cos(\alpha_{\text{el}}),\ 
  \sin(\alpha_{\text{el}})
\end{bmatrix}^\mathsf{T} .
\end{equation}

\subsection{Geometry Mismatch}
Taking the \ac{bs}'s position as a reference, we can focus on the calibration errors in $\mathbf{p}_\text{r}$ and $\mathbf{R}_\text{r}$. Thus, the accessible prior geometry information of the \ac{ris} can be modeled as 
\begin{align}
  \tilde{\mathbf{p}}_\text{r} = {\mathbf{p}}_\text{r} + \mathbf{u},\qquad
  \tilde{\mathbf{R}}_\text{r} = {\mathbf{R}}_\text{p}(\mathbf{v}){\mathbf{R}}_\text{r},
\end{align}
where $\mathbf{u}$
and $\mathbf{v}$ are the calibration errors that cause the mismatch in the model.
Here, ${\mathbf{R}}_\text{p}(\mathbf{v})= \mathbf{R}_z(v_3)\mathbf{R}_y(v_2)\mathbf{R}_x(v_1)$,
where $\mathbf{R}_x(v_1)$ denotes a rotation of $v_1$ degree around the X-axis, and likewise for $\mathbf{R}_y(v_2)$ and $\mathbf{R}_z(v_3)$.

\section{Localization Lower Bound}
\subsection{Misspecified Cram\'er-Rao Bound}
Suppose a two-stage localization framework is employed~\cite{Shahmansoori2018Position,chen2022tutorial}, that consists of a channel parameter estimation process followed by a UE location estimation from the obtained channel parameters.
For the channel parameter estimation stage, we estimate the following channel parameters from the received signal $\mathbf{y}_\ell, \ell=1,\dots,L$ given in \eqref{eq_channel}:
\begin{equation}\label{eq_eta_ch}
    \bm{\eta}_{\text{ch}} \triangleq [ {\phi}_{\text{az}}, {\phi}_{\text{el}}, \tau_\text{b}, {\tau}_\text{r},\mathfrak{R}(g_\text{b}),\mathfrak{I}(g_\text{b}),\mathfrak{R}(g_\text{r}),\mathfrak{I}(g_\text{r}) ]^\mathsf{T},
\end{equation}
where $\mathfrak{R}(\cdot) $ / $\mathfrak{I}(\cdot)$ denote the operations of taking the real / imaginary part.
The \ac{fim} of $\bm{\eta}_\text{ch}$ can be calculated by the Slepian-Bangs formula~\cite[Sec. 3.4]{kay1993fundamentals}
\vspace{-0.7em}
\begin{equation}\label{eq_FIMeta}
    \mathbf{J}({\bm{\eta}_\text{ch}}) = \frac{2}{\sigma^2}\sum_{\ell = 1}^{L} \mathfrak{R}\left( \left(\frac{\partial \bm{\mu}_\ell}{\partial \bm{\eta}_\text{ch}}\right)^\mathsf{H}  \frac{\partial \bm{\mu}_\ell}{\partial \bm{\eta}_\text{ch}}\right).
    \vspace{-0.2em}
\end{equation}
By removing the nuisance parameters related to $g_\text{b}$ and $g_\text{r}$, we further define $\bm{\eta}\triangleq[{\phi}_{\text{az}}, {\phi}_{\text{el}}, \tau_\text{b}, {\tau}_\text{r}]^\mathsf{T} $ used for position and orientation estimation.
We can compute the \ac{fim} for $\bm{\eta}$ using Schur's complement: we partition $\mathbf{J}(\bm{\eta}_\text{ch})=[\mathbf{X},\mathbf{Y};\mathbf{Y}^\mathsf{T},\mathbf{Z}]$,
where $\mathbf{X}\in\mathbb{R}^{4\times 4}$ so that $\mathbf{J}(\bm{\eta})=\mathbf{X}-\mathbf{Y}\mathbf{Z}^{-1}\mathbf{Y}^\mathsf{T}$.

For the \ac{ue} location estimation, 
we define a parameter vector in the location domain as 
$
    \mathbf{r} \triangleq [ \mathbf{p}^\mathsf{T}, \Delta ]^\mathsf{T}\in\mathbb{R}^4.
$
The objective here is to estimate $\mathbf{r}$ from the estimated $\hat{\bm{\eta}}$, where the \ac{ris} geometry mismatch occurs.
We express $\bm{\eta}$ as a function of $\mathbf{r}$, i.e., $\bm{\eta} = \mathbf{g}(\mathbf{r}|\mathbf{p}_\text{b},\mathbf{p}_\text{r},\mathbf{R}_\text{r})$, a relationship that is defined by~\eqref{eq_phi}-\eqref{eq_tau}.
Thus, the distribution of the estimated $\hat{\bm{\eta}}$ can be obtained as $\hat{\bm{\eta}}\sim\mathcal{N}\left(\mathbf{g}(\mathbf{r}|\mathbf{p}_\text{b},\mathbf{p}_\text{r},\mathbf{R}_\text{r}),\bm{\Sigma}\right)$, where $\bm{\Sigma} = \mathbf{J}^{-1}(\bm{\eta})$ if an \emph{efficient} channel estimator is applied in the first stage.
To estimate $\mathbf{r}$, however, we adopt the mismatched model by using $\hat{\bm{\eta}}\sim\mathcal{N}\left(\mathbf{g}({\mathbf{r}}|\mathbf{p}_\text{b},\tilde{\mathbf{p}}_\text{r},\tilde{\mathbf{R}}_\text{r}),\bm{\Sigma}\right)$ to obtain an estimate $\hat{\mathbf{r}}$ from $\hat{\bm{\eta}}$.
Then, we have the true likelihood function ($f_\text{T}$) and the mismatched likelihood function ($f_\text{M}$) as
\begin{align}\notag
    \ln f_\text{T} &= -\frac{1}{2}(\hat{\bm{\eta}}-\mathbf{g}(\mathbf{r}|\mathbf{p}_\text{r},\mathbf{R}_\text{r}))^\mathsf{T} \bm{\Sigma}^{-1} (\hat{\bm{\eta}}-\mathbf{g}(\mathbf{r}|\mathbf{p}_\text{r},\mathbf{R}_\text{r})),\\
    \ln f_\text{M} &= -\frac{1}{2}(\hat{\bm{\eta}}-\mathbf{g}(\mathbf{r}|\tilde{\mathbf{p}}_\text{r},\tilde{\mathbf{R}}_\text{r}))^\mathsf{T} \bm{\Sigma}^{-1} (\hat{\bm{\eta}}-\mathbf{g}(\mathbf{r}|\tilde{\mathbf{p}}_\text{r},\tilde{\mathbf{R}}_\text{r})).\notag
\end{align}
Note that we omit the constant term and the parameter $\mathbf{p}_\text{b}$ as they do not affect the estimation.

Now, the lower bound matrix of the estimation \ac{mse}  based on $f_\text{M}$ can be obtained as~\cite{chen2022channel}
\begin{equation}\label{eq_MCRB_matrix}
    \text{LBM}(\hat{\mathbf{r}},\bar{\mathbf{r}}) = \underbrace{\mathbf{A}_{\mathbf{r}_0}^{-1} \mathbf{B}_{\mathbf{r}_0} \mathbf{A}_{\mathbf{r}_0}^{-1}}_{\text{MCRB}(\mathbf{r}_0)} + \underbrace{(\bar{\mathbf{r}}-\mathbf{r}_0)(\bar{\mathbf{r}}-\mathbf{r}_0)^\mathsf{T}}_{\text{Bias}(\mathbf{r}_0)},
\end{equation}
where $\bar{\mathbf{r}}$ is the true parameter vector, $\mathbf{r}_0$ is the pseudo-true parameter vector that minimizes the \ac{kld} between $f_\text{T}$ and $f_\text{M}$. $\mathbf{A}_{\mathbf{r}_0}$ and $\mathbf{B}_{\mathbf{r}_0}$ are two generalizations of the FIMs as~\cite{fortunati2017performance}
\begin{equation}\label{eq_KLD}
    \mathbf{r}_0 = \argmin_\mathbf{r} D(f_\text{T}(\hat{\bm{\eta}}|\bar{\mathbf{r}})\|f_\text{M}(\hat{\bm{\eta}}|\mathbf{r})),
\end{equation}
\vspace{-2em}
\begin{align}
  \left[\mathbf{A}_{\mathbf{r}_0}\right]_{i,j} &= \mathbb{E}_{f_\text{T}}\left\{\frac{\partial^2}{\partial r_i\partial r_j}\ln f_\text{M}(\hat{\bm{\eta}}|\mathbf{r})\Big|_{\mathbf{r}=\mathbf{r}_0} \right\},\\
  &= \left(\frac{\partial^2\mathbf{g}(\mathbf{r}|\tilde{\mathbf{p}}_\text{r},\tilde{\mathbf{o}}_\text{r})}{\partial r_i\partial r_j}\right)^\mathsf{T} 
  \bm{\Sigma}^{-1}(\bm{\eta}-\mathbf{g}(\mathbf{r}|\tilde{\mathbf{p}}_\text{r},\tilde{\mathbf{o}}_\text{r}))\Big|_{\mathbf{r}=\mathbf{r}_0} \notag\\
  &\qquad\quad -\left(\frac{\partial\mathbf{g}(\mathbf{r}|\tilde{\mathbf{p}}_\text{r},\tilde{\mathbf{o}}_\text{r})}{\partial r_i}\right)^\mathsf{T}  \bm{\Sigma}^{-1}\frac{\partial\mathbf{g}(\mathbf{r}|\tilde{\mathbf{p}}_\text{r},\tilde{\mathbf{o}}_\text{r})}{\partial r_j}\bigg|_{\mathbf{r}=\mathbf{r}_0},\notag
\end{align}
\vspace{-2em}
\begin{align}
  \mathbf{B}_{\mathbf{r}_0} &= \mathbb{E}_{f_\text{T}}\left\{  
  \frac{\partial\ln f_\text{M}(\hat{\bm{\eta}}|\mathbf{r})}{\partial \mathbf{r}}\Big|_{\mathbf{r}=\mathbf{r}_0}
  \cdot\left(\frac{\partial\ln f_\text{M}(\hat{\bm{\eta}}|\mathbf{r})}{\partial \mathbf{r}}\Big|_{\mathbf{r}=\mathbf{r}_0}\right)^\mathsf{T}
  \right\},\notag\\
  &=\left(\frac{\partial\mathbf{g}(\mathbf{r}|\tilde{\mathbf{p}}_\text{r},\tilde{\mathbf{o}}_\text{r})}{\partial \mathbf{r}}\right)^\mathsf{T} \bm{\Sigma}^{-1}
  \tilde{\bm{\Sigma}}(\mathbf{r})
  \bm{\Sigma}^{-1}  \frac{\partial\mathbf{g}(\mathbf{r}|\tilde{\mathbf{p}}_\text{r},\tilde{\mathbf{o}}_\text{r})}{\partial \mathbf{r}}  \bigg|_{\mathbf{r}=\mathbf{r}_0}\notag,
\end{align}
where $D(\cdot\|\cdot)$ denotes the \ac{kld} and $\tilde{\bm{\Sigma}}(\mathbf{r})=\bm{\Sigma}+(\bm{\eta}-\mathbf{g}(\mathbf{r}|\tilde{\mathbf{p}}_\text{r},\tilde{\mathbf{o}}_\text{r}))  (\bm{\eta}-\mathbf{g}(\mathbf{r}|\tilde{\mathbf{p}}_\text{r},\tilde{\mathbf{o}}_\text{r}))^\mathsf{T}$.
Therefore, the remaining problem is solving \eqref{eq_KLD} to obtain the pseudo-true parameter $\mathbf{r}_0$.
As mentioned, most previous works on localization mismatch analysis solve \eqref{eq_KLD} using numerical methods, e.g., gradient descent, which is time-consuming and offers no guarantees to obtain the global minimum.
The next subsection presents a closed-form solution to the pseudo-true parameter estimation problem with the global minimum guarantee.

\subsection{The Closed-form Pseudo-true Parameter Vector}\label{sec_closedform}
In this subsection, we derive a closed-form solution for \eqref{eq_KLD}.
According to the definition of \ac{kld}, we have 
\begin{align}
  &D(f_\text{T}(\hat{\bm{\eta}}|\bar{\mathbf{r}})\|f_\text{M}(\hat{\bm{\eta}}|\mathbf{r})) = \mathbb{E}_{f_\text{T}}\left\{\ln f_\text{T}(\hat{\bm{\eta}}|\bar{\mathbf{r}}) - \ln f_\text{M}(\hat{\bm{\eta}}|\mathbf{r})\right\},\notag\\
  &= -\frac{1}{2}\mathbb{E}_{f_\text{T}}\left\{ (\hat{\bm{\eta}}-\mathbf{g}(\bar{\mathbf{r}}|\mathbf{p}_\text{r},\mathbf{R}_\text{r}))^\mathsf{T} \bm{\Sigma}^{-1} (\hat{\bm{\eta}}-\mathbf{g}(\bar{\mathbf{r}}|\mathbf{p}_\text{r},\mathbf{R}_\text{r}))\right\}\notag \\
&\qquad + \frac{1}{2}\mathbb{E}_{f_\text{T}}\left\{(\hat{\bm{\eta}}-\mathbf{g}(\mathbf{r}|\tilde{\mathbf{p}}_\text{r},\tilde{\mathbf{R}}_\text{r}))^\mathsf{T} \bm{\Sigma}^{-1} (\hat{\bm{\eta}}-\mathbf{g}(\mathbf{r}|\tilde{\mathbf{p}}_\text{r},\tilde{\mathbf{R}}_\text{r}))\right\},\notag \\
&\stackrel{(a)}{=}\frac{1}{2}\mathbf{h}(\mathbf{r})^\mathsf{T} \bm{\Sigma}^{-1} \mathbf{h}(\mathbf{r}),\label{eq_quad}
\end{align}
where $\mathbf{h}(\mathbf{r})\triangleq\mathbf{g}(\bar{\mathbf{r}}|\mathbf{p}_\text{r},\mathbf{R}_\text{r})-\mathbf{g}(\mathbf{r}|\tilde{\mathbf{p}}_\text{r},\tilde{\mathbf{R}}_\text{r})$ 
and step $(a)$ can be obtained by using~\cite[Eq. 380]{petersen2008matrix}.
Since $\bm{\Sigma}$ is a covariance matrix, then $\bm{\Sigma}^{-1}$ is positive definite. 
Hence, the quadratic form \eqref{eq_quad} always satisfies
\begin{equation}
  D(f_\text{T}(\hat{\bm{\eta}}|\bar{\mathbf{r}})\|f_\text{M}(\hat{\bm{\eta}}|\mathbf{r})) = \frac{1}{2}\mathbf{h}(\mathbf{r})^\mathsf{T} \bm{\Sigma}^{-1} \mathbf{h}(\mathbf{r}) \geqslant 0, \label{eq_minimum}
\end{equation}
where the equality holds if and only if $\mathbf{h}(\mathbf{r})=\mathbf{0}$.
The inequality in \eqref{eq_minimum} implies that if there exists a vector $\mathbf{r}_0$ that satisfies $\mathbf{h}(\mathbf{r}_0)=\mathbf{0}$, then it must be the global minimum of~\eqref{eq_KLD}.
The following steps aim to find such a vector $\mathbf{r}_0$, which is possible in the considered problem since the number of unknowns ($\mathbf{r}\in\mathbb{R}^4$) equals the number of observations ($\mathbf{h}(\mathbf{r})\in\mathbb{R}^4$).

Let $\mathbf{r}_0 = [{\mathbf{p}}^\mathsf{T}_0, {\Delta}_0]^\mathsf{T}$ be the pseudo-true parameters and $[\bar{\mathbf{p}}^\mathsf{T}, \bar{\Delta}]^\mathsf{T} =\bar{\mathbf{r}}$, then $\mathbf{h}(\mathbf{r}_0)=\mathbf{0}$ can be rewritten based on \eqref{eq_phi}--\eqref{eq_tau} as
\begin{align}\label{eq_1}
  &{\|\mathbf{p}_\text{b}-\bar{\mathbf{p}}\|} + {c}\bar{\Delta} = {\|\mathbf{p}_\text{b}-\mathbf{p}_0\|} + {c}\Delta_0 , \\
  &{\|\mathbf{p}_\text{b}-\mathbf{p}_\text{r}\| + \|\mathbf{p}_\text{r}-\bar{\mathbf{p}}\|} + {c}\bar{\Delta} = \|\mathbf{p}_\text{b}-\tilde{\mathbf{p}}_\text{r}\|\notag \\ 
    &\qquad\qquad\qquad\qquad\qquad\qquad + \|\tilde{\mathbf{p}}_\text{r}-\mathbf{p}_0\| + {c}\Delta_0,\\
  & {\mathbf{R}_\text{r}^\mathsf{T} (\bar{\mathbf{p}}-\mathbf{p}_\text{r})}/{\|\bar{\mathbf{p}}-\mathbf{p}_\text{r}\|} = {\tilde{\mathbf{R}}_\text{r}^\mathsf{T} (\mathbf{p}_0-\tilde{\mathbf{p}}_\text{r})}/{\|\mathbf{p}_0-\tilde{\mathbf{p}}_\text{r}\|}.\label{eq_3}
\end{align}
Combining \eqref{eq_1}--\eqref{eq_3}, we obtain 
\begin{equation}\label{eq_intersec}
  (\alpha + \|\mathbf{p}_0-\tilde{\mathbf{p}}_\text{r}\| - \|\mathbf{p}_0-{\mathbf{p}}_\text{b}\|)\frac{\mathbf{p}_0-\tilde{\mathbf{p}}_\text{r}}{\|\mathbf{p}_0-\tilde{\mathbf{p}}_\text{r}\|} = \tilde{\mathbf{R}}_\text{r}\mathbf{R}_\text{r}^\mathsf{T}(\bar{\mathbf{p}}-\mathbf{p}_\text{r}),
\end{equation}
where $\alpha=\|{\mathbf{p}}_\text{b}-\tilde{\mathbf{p}}_\text{r}\| + \|{\mathbf{p}}_\text{b}-\bar{\mathbf{p}}\| - \|{\mathbf{p}}_\text{b}-{\mathbf{p}}_\text{r}\|$.
From \eqref{eq_intersec}, it can be inferred that a pseudo-true \ac{ue} position can be obtained as the intersection of a line $s_l$ and a hyperboloid $s_h$ given by
\begin{align}
  s_l:\quad & \mathbf{p} = x\tilde{\mathbf{R}}_\text{r}\mathbf{R}_\text{r}^\mathsf{T}(\bar{\mathbf{p}}-\mathbf{p}_\text{r}) + \tilde{\mathbf{p}}_\text{r},\label{eq_line}\\
  s_h:\quad & \|{\mathbf{p}}-\tilde{\mathbf{p}}_\text{r}\| - \|{\mathbf{p}}-{\mathbf{p}}_\text{b}\| = \beta,\label{eq_hyperboloid}
\end{align}
where $x$ is a positive scalar representing the length of the line segment, and $\beta=\|{\bar{\mathbf{p}}}-{\mathbf{p}}_\text{r}\|-\alpha$.
As a result, the pseudo-true \ac{ue} position can be determined by solving $x$, which is
\begin{equation}
  x_0 = \frac{\beta^2-\|\tilde{\mathbf{p}}_\text{r}-\mathbf{p}_\text{b}\|^2}{2[\mathbf{a}^\mathsf{T}(\tilde{\mathbf{p}}_\text{r}-\mathbf{p}_\text{b}) + \beta\|\mathbf{a}\| ]},
\end{equation}
with $\mathbf{a}=\tilde{\mathbf{R}}_\text{r}\mathbf{R}_\text{r}^\mathsf{T}(\bar{\mathbf{p}}-\mathbf{p}_\text{r})$. Then, the pseudo-true \ac{ue} position ($\mathbf{p}_0$) is obtained by substituting $x_0$ into~\eqref{eq_line}.
An estimate of pseudo-true clock bias $\Delta_0$ can be directly determined from \eqref{eq_1} given $\mathbf{p}_0$.

\subsection{Mismatched Bound and Estimator}

Based on \eqref{eq_MCRB_matrix}, the expected \ac{rmse} of the \ac{ue} position estimation (under \ac{ris} geometry mismatch) can be lower bounded as 
\begin{equation}\label{eq_LB}
   \sqrt{\mathbb{E}\{\|\mathbf{p}-\hat{\mathbf{p}}\|^2\}} \geq \sqrt{\text{tr}\big(\left[\text{LBM}(\hat{\mathbf{r}},\bar{\mathbf{r}})\right]_{1:3,1:3}\big)} \triangleq \text{LB},
\end{equation}
where $\text{tr}(\cdot)$ returns the trace of a matrix.

To verify the derived \ac{mcrb}, we formulate a \ac{ml}-based localization estimator.
To simplify the analysis, we assume that an efficient estimator is applied to obtain the channel parameters $\hat{\bm{\eta}}$. As such, we focus on analyzing the \ac{ue} position estimation stage where the mismatch impact appears.
Hence we have the following misspecified maximum likelihood (MML) estimation:
\begin{equation}\label{eq_ML}
    \hat{\mathbf{r}}_\text{MML} = \argmax_{\mathbf{r}}\  \ln f_\text{M}(\hat{\bm{\eta}}_\text{ch}|\mathbf{r}),
\end{equation}
which can be solved using the gradient descent method.

\section{Simulation Results}
\subsection{Simulation Setup}
We consider a localization system that consists of a \ac{bs} at $[5,0,3]^\mathsf{T}$, 
a \ac{ris} at $[0,-5,2.5]^\mathsf{T}$ with orientation $\mathbf{o}_\text{r}=[0^\circ,0^\circ,90^\circ]$, 
and a \ac{ue} with the default location $[-2.5,2.5,0]^\mathsf{T} $.
The \ac{ris} is of size $64\times 64$ elements with half-wavelength spacing. The other parameters are as follows: average transmission power $P = \unit[10]{dBm}$, carrier frequency $f_c = \unit[28]{GHz}$, bandwidth $W = \unit[400]{MHz}$, number of transmissions $L = 32$, number of subcarriers $K = 3000$, noise PSD $N_0 = \unit[-173.855]{dBm/Hz}$ and noise figure $N_f = \unit[10]{dB}$.
\vspace{-1em}
\subsection{Results Analysis}
\begin{figure}[t]
  \centering
  \includegraphics[width=3.35in]{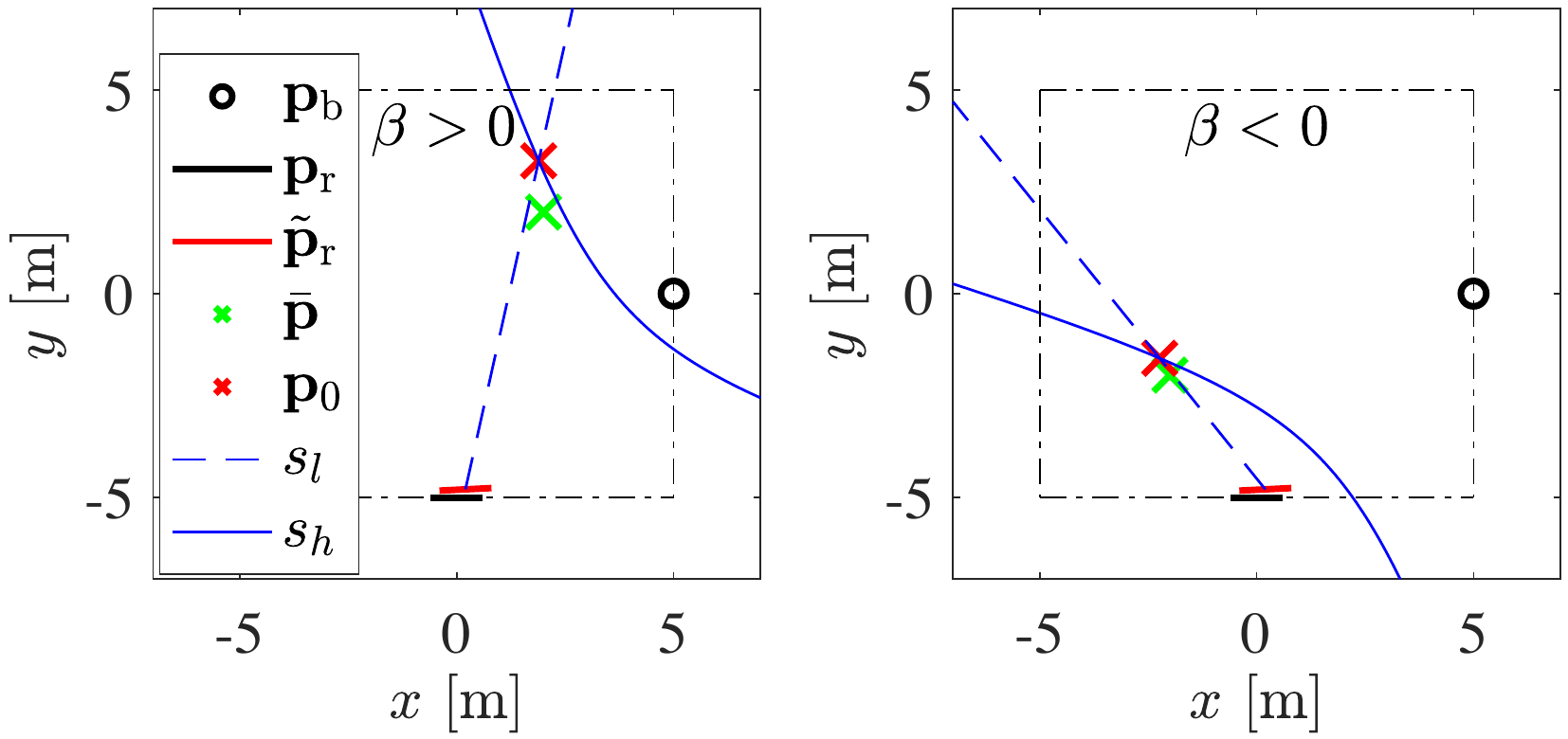}
  \vspace{-5mm}
  \caption{ 
    The geometric relationship between the pseudo-true \ac{ue} position $\mathbf{p}_0$ and true \ac{ue} position $\bar{\mathbf{p}}$.
 }
  \label{fig_pseudo}
  \vspace{-3mm}
\end{figure}

We first present the geometry relationship between the pseudo-true \ac{ue} position $\mathbf{p}_0$ and the true \ac{ue} position $\bar{\mathbf{p}}$. 
We set a fixed \ac{ris} position and orientation mismatch error as \unit[$\mathbf{u}=0.2\times\mathbf{1}$]{m} and \unit[$\mathbf{v}=3\times\mathbf{1}$]{deg}, respectively.
Fig.~\ref{fig_pseudo} shows the projection of the corresponding geometry onto the XOY plane.
We can see that the \ac{ris} geometry mismatch causes a deviation between the true and pseudo-true \ac{ue} position, 
and the pseudo-true \ac{ue} position $\mathbf{p}_0$ (obtained by solving~\eqref{eq_KLD} with a gradient descent method) coincides with the intersection of the line $s_l$ and hyperboloid $s_h$,
which affirms the solution stated in Subsection~\ref{sec_closedform}.
We can also observe that depending on the position of the \ac{ue}, either $\beta>0$ or $\beta<0$ can occur. An example of each case is shown in Fig.~\ref{fig_pseudo}.

\begin{figure}[t]
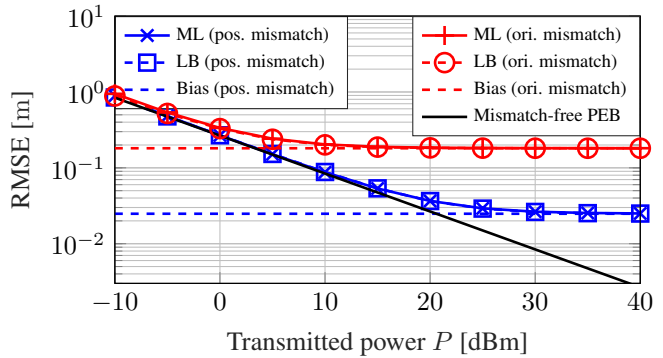

    \begin{minipage}[b]{0.1\linewidth}
      \centering
        \include{figures/MCRB_vs_P}
        \vspace{-0.8cm}
    \end{minipage}
    \caption{
    ML-RMSE, LB, and bias term versus transmitted power (SNR) for RIS position mismatch (\unit[$\mathbf{u}=0.01\times\mathbf{1}$]{m}) and orientation mismatch (\unit[$\mathbf{v}=0.5\times\mathbf{1}$]{deg}).
    }
    \label{fig_MCRB_vs_P}
    \vspace{-3mm}
\end{figure}

Next, we evaluate the position estimation performance of the estimator \eqref{eq_ML} and compare it with the theoretical bound  LB \eqref{eq_LB} to validate the derived pseudo-true parameter vector $\mathbf{r}_0$ and the \ac{mcrb}.
Fig.~\ref{fig_MCRB_vs_P} shows the \ac{rmse} for the \ac{ml} estimator~\cite{manopt}, LB, bias term, and the mismatch-free \ac{peb} versus different transmitted power for a \ac{ris} position mismatch (\unit[$\mathbf{u}=0.01\times\mathbf{1}$]{m}) and orientation mismatch (\unit[$\mathbf{v}=0.5\times\mathbf{1}$]{deg}) separately.
We can observe that at low transmit power (i.e., low \ac{snr} given a fixed noise) levels, the LB and the mismatch-free \ac{peb} coincide, implying that the \ac{ris} geometry mismatch is not the main source of error. 
At higher signal \ac{snr}, however, LB deviates from the mismatch-free \ac{peb} and saturates, which reveals the positioning performance is thus more severely affected by the \ac{ris} geometry mismatch.
The \ac{rmse} of the \ac{ml} estimator closely follows the LB, which demonstrates the validity of our derivation.

\begin{figure}[t]
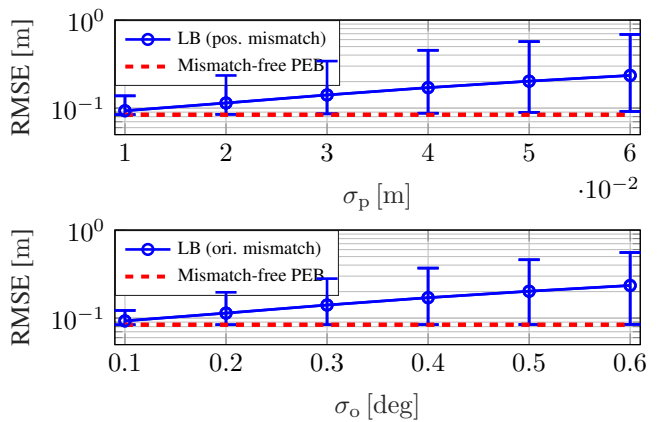

    \begin{minipage}[b]{0.1\linewidth}
      \centering
        \include{figures/MCRB_vs_sigma2}
        \vspace{-0.8cm}
    \end{minipage}
    \caption{
      LB and PEB versus different levels of \ac{ris} position mismatch ($\sigma_\text{p}=\unit[\{1,2,3,4,5,6\}\times 10^{-2}]{m}$) and orientation mismatch ($\sigma_\text{o}=\unit[\{1,2,3,4,5,6\}\times 10^{-1}]{deg}$).
    }
    \label{fig_MCRB_vs_sigma2}
    \vspace{-3mm}
\end{figure}
Finally, in Fig.\ref{fig_MCRB_vs_sigma2}, we evaluate the localization performance as a function of the standard deviation of the mismatch errors. 
We assume $\mathbf{u}\sim\mathcal{N}(\mathbf{0},\sigma_{\text{p}}^2\mathbf{I}_3)$, $\mathbf{v}\sim\mathcal{N}(\mathbf{0},\sigma_{\text{o}}^2\mathbf{I}_3)$, and generate 100 statistical realizations of the mismatch for each standard deviation value and collect the minimum, maximum, and mean values of the LBs.
It can be seen that, for both position and orientation mismatches, a larger mismatch standard deviation results in an average higher performance bound and produces a larger perturbation in the LB.

\section{Conclusion}
This paper considered a \ac{siso} \ac{ris}-aided localization system with \ac{ris} geometry mismatch.
We proposed a closed-form solution to determine the pseudo-true parameters, which is used for \ac{mcrb} derivation.
The derived \ac{mcrb} is validated using the empirical \ac{rmse} of a \ac{ml} estimator. A phenomenon is observed whereby the bound saturates as the \ac{snr} increases. 

\clearpage

\bibliographystyle{IEEEbib}
\bibliography{reference}

\end{document}

%% file: Acronyms.tex
\DeclareAcronym{5g}{
short=5G,
long= fifth generation,
}
\DeclareAcronym{6g}{
short=6G,
long= sixth generation,
}
\DeclareAcronym{3d}{
short=3D,
long= three-dimensional,
}
\DeclareAcronym{2d}{
short=2D,
long= two-dimensional,
}
\DeclareAcronym{1d}{
short=1D,
long= one-dimensional,
}
\DeclareAcronym{aod}{
short=AOD,
long= angle-of-departure,
}
\DeclareAcronym{aosa}{
short=AOSA,
long= array-of-subarray,
}
\DeclareAcronym{adod}{
short=ADOD,
long= angle-difference-of-departure,
}
\DeclareAcronym{aoa}{
short=AOA,
long= angle-of-arrival,
}
\DeclareAcronym{adc}{
short=ADC,
long= analog to digital converter,
}
\DeclareAcronym{aeb}{
short=AEB,
long= angle error bound,
}
\DeclareAcronym{av}{
short=AV,
long= autonomous vehicle,
}
\DeclareAcronym{bs}{
short=BS,
long= base station,
}

\DeclareAcronym{csi}{
short=CSI,
long= channel state information,
}
\DeclareAcronym{cfo}{
short=CFO,
long= carrier frequency offset,
}
\DeclareAcronym{ceb}{
short=CEB,
long= clock error bound,
}
\DeclareAcronym{coa}{
short=COA,
long= curvature-of-arrival,
}
\DeclareAcronym{crb}{
short=CRB,
long= Cram\'er-Rao bound,
}
\DeclareAcronym{ccrb}{
short=CCRB,
long= constrained Cram\'er-Rao bound,
}
\DeclareAcronym{cmos}{
short=CMOS,
long= complementary metal-oxide-semiconductor,
}
\DeclareAcronym{crlb}{
short=CRLB,
long= Cram\'er-Rao lower bound,
}
\DeclareAcronym{cdf}{
short=CDF,
long= cumulative distribution function,
}
\DeclareAcronym{cp}{
short=CP,
long= cyclic prefix,
}
\DeclareAcronym{dac}{
short=DAC,
long= digital to analog converter,
}
\DeclareAcronym{dfl}{
short=DFL,
long= device-free localization,
}
\DeclareAcronym{dmimo}{
short=D-MIMO,
long= distributed MIMO,
}
\DeclareAcronym{dlprs}{
short=DL-PRS,
long= downlink positioning reference signal,
}
\DeclareAcronym{d2d}{
short=D2D,
long= device-to-device,
}
\DeclareAcronym{dftsofdm}{
short=DFT-s-OFDM,
long= discrete-Fourier-transform spread OFDM,
}
\DeclareAcronym{dl}{
short=DL,
long= deep learning,
}
\DeclareAcronym{gps}{
short=GPS,
long= global positioning system,
}
\DeclareAcronym{gnss}{
short=GNSS,
long= global navigation satellite system,
}
\DeclareAcronym{fim}{
short=FIM,
long = Fisher information matrix,
}
\DeclareAcronym{hwi}{
short=HWI,
long= hardware impairment,
}
\DeclareAcronym{hemt}{
short=HEMT,
long= high electron mobility transistor,
}
\DeclareAcronym{hbt}{
short=HBT,
long= heterojunction bipolar transistors,
}
\DeclareAcronym{iot}{
short=IoT,
long= internet of things,
}
\DeclareAcronym{imu}{
short=IMU,
long= inertial measurement unit,
}
\DeclareAcronym{isac}{
short=ISAC,
long= integrated sensing and communication,
}
\DeclareAcronym{iqi}{
short=IQI,
long= in-phase and quadrature imbalance,
}
\DeclareAcronym{ip}{
short=IP,
long= incidence point,
}
\DeclareAcronym{ia}{
short=IA,
long= initial access,
}
\DeclareAcronym{kpi}{
short=KPI,
long= key performance indicator,
}
\DeclareAcronym{kf}{
short=KF,
long= Kalman filter,
}
\DeclareAcronym{ekf}{
short=EKF,
long= extended Kalman filter,
}
\DeclareAcronym{ukf}{
short=UKF,
long= unscented Kalman filter,
}
\DeclareAcronym{ckf}{
short=CKF,
long= cubature Kalman filter,
}
\DeclareAcronym{pf}{
short=PF,
long= particle filter,
}
\DeclareAcronym{lb}{
short=LB,
long= lower bound,
}
\DeclareAcronym{lse}{
short=LSE,
long= least-square estimation,
}
\DeclareAcronym{ls}{
short=LS,
long= least-squares,
}
\DeclareAcronym{lo}{
short=LO,
long= local oscillator,
}
\DeclareAcronym{los}{
short=LOS,
long= line-of-sight,
}
\DeclareAcronym{mc}{
short=MC,
long= mutual coupling,
}
\DeclareAcronym{mac}{
short=MAC,
long= medium access control,
}
\DeclareAcronym{meb}{
short=MEB,
long= mapping error bound,
}
\DeclareAcronym{mcrb}{
short=MCRB,
long= misspecified Cram\'er-Rao bound,
}
\DeclareAcronym{mds}{
short=MDS,
long= multidimensional scaling ,
}
\DeclareAcronym{mimo}{
short=MIMO,
long= multiple-input-multiple-output,
}
\DeclareAcronym{siso}{
short=SISO,
long= single-input-single-output,
}
\DeclareAcronym{mm}{
short=MM,
long= mismatched model,
}
\DeclareAcronym{mpc}{
short=MPC,
long= multipath components,
}
\DeclareAcronym{mmwave}{
short=mmWave,
long= millimeter wave,
}
\DeclareAcronym{mmle}{
short=MMLE,
long= mismatched maximum likelihood estimation,
}
\DeclareAcronym{mems}{
short=MEMS,
long= micro-electro-mechanical system,
}
\DeclareAcronym{mle}{
short=MLE,
long= maximum likelihood estimation,
}
\DeclareAcronym{ml}{
short=ML,
long= maximum likelihood,
}
\DeclareAcronym{nlos}{
short=NLOS,
long= none-line-of-sight,
}
\DeclareAcronym{ofdm}{
short=OFDM,
long= orthogonal frequency-division multiplexing,
}
\DeclareAcronym{oeb}{
short=OEB,
long= orientation error bound,
}
\DeclareAcronym{otfs}{
short=OTFS,
long= orthogonal time-frequency space,
}
\DeclareAcronym{pdf}{
short=PDF,
long= probability density function,
}
\DeclareAcronym{papr}{
short=PAPR,
long= peak-to-average-power ratio,
}
\DeclareAcronym{pan}{
short=PAN,
long= power amplifier nonlinearity,
}
\DeclareAcronym{pa}{
short=PA,
long= power amplifier,
}
\DeclareAcronym{ps}{
short=PS,
long= phase shifter,
}
\DeclareAcronym{pn}{
short=PN,
long= phase noise,
}
\DeclareAcronym{poa}{
short=POA,
long= phase-of-arrival,
}
\DeclareAcronym{pwm}{
short=PWM,
long= planar wave model,
}
\DeclareAcronym{pdoa}{
short=PDOA,
long= phase-difference-of-arrival,
}
\DeclareAcronym{prs}{
short=PRS,
long= positioning reference signals,
}
\DeclareAcronym{peb}{
short=PEB,
long= position error bound,
}
\DeclareAcronym{speb}{
short=SPEB,
long= squared position error bound,
}
\DeclareAcronym{rnn}{
short=RNN,
long= recurrent neural network,
}
\DeclareAcronym{rl}{
short=RL,
long= reinforcement learning,
}
\DeclareAcronym{rfc}{
short=RFC,
long= radio-frequency chain,
}
\DeclareAcronym{rf}{
short=RF,
long= radio frequency,
}
\DeclareAcronym{rfid}{
short=RFID,
long= radio frequency identification,
}
\DeclareAcronym{ris}{
short=RIS,
long= reconfigurable intelligent surface,
}
\DeclareAcronym{rss}{
short=RSS,
long= received signal strength,
}
\DeclareAcronym{rtt}{
short=RTT,
long= round-trip time,
}
\DeclareAcronym{sm}{
short=SM,
long= standard model,
}
\DeclareAcronym{sige}{
short=SiGe,
long= silicon-germanium,
}
\DeclareAcronym{spp}{
short=SPP,
long= surface plasmon polariton,
}
\DeclareAcronym{sa}{
short=SA,
long= subarray,
}
\DeclareAcronym{sota}{
short=SOTA,
long= state-of-the-art,
}
\DeclareAcronym{swm}{
short=SWM,
long= spherical wave model,
}
\DeclareAcronym{slam}{
short=SLAM,
long= simultaneous localization and mapping,
}
\DeclareAcronym{tm}{
short=TM,
long= true model,
}
\DeclareAcronym{toa}{
short=TOA,
long= time-of-arrival,
}
\DeclareAcronym{tof}{
short=TOF,
long= time-of-flight,
}
\DeclareAcronym{tdoa}{
short=TDOA,
long= time-difference-of-arrival,
}
\DeclareAcronym{thz}{
short=THz,
long= Terahertz,
}
\DeclareAcronym{ue}{
short=UE,
long= user equipment,
}
\DeclareAcronym{ummimo}{
short=UM-MIMO,
long= ultra-massive multi-input-multi-output,
}
\DeclareAcronym{vlp}{
short=VLP,
long= visible light positioning,
}
\DeclareAcronym{veb}{
short=VEB,
long= velocity error bound,
}
\DeclareAcronym{vlc}{
short=VLC,
long= visible light communication,
}
\DeclareAcronym{ula}{
short=ULA,
long= uniform linear array,
}
\DeclareAcronym{uav}{
short=UAV,
long= unmanned aerial vehicle,
}
\DeclareAcronym{upa}{
short=UPA,
long= uniform planar array,
}
\DeclareAcronym{wlan}{
short=WLAN,
long= wireless local area network,
}
\DeclareAcronym{gcs}{
short=GCS,
long= global coordinate system,
}
\DeclareAcronym{lcs}{
short=LCS,
long= local coordinate system,
}
\DeclareAcronym{awgn}{
short=AWGN,
long= additive white Gaussian noise,
}
\DeclareAcronym{iid}{
short=i.i.d.,
long= independent and identically distributed,
}
\DeclareAcronym{snr}{
short=SNR,
long= signal-to-noise ratio,
}
\DeclareAcronym{csir}{
short=CSIR,
long= channel state information at the receiver,
}
\DeclareAcronym{cdit}{
short=CDIT,
long= channel distribution information at the transmitter,
}
\DeclareAcronym{tbps}{
short=Tbps,
long= Terabit-per-second,
}
\DeclareAcronym{um-mimo}{
short=UM-MIMO,
long= ultra-massive multiple-input multiple-output,
}
\DeclareAcronym{rmse}{
short=RMSE,
long= root mean square error,
}
\DeclareAcronym{mse}{
short=MSE,
long= mean squared error,
}
\DeclareAcronym{ccdf}{
short=CCDF,
long= complementary cumulative distribution function,
}
\DeclareAcronym{ae}{
short=AE,
long= antenna element,
}
\DeclareAcronym{mrc}{
short=MRC,
long= maximum ratio combining,
}
\DeclareAcronym{egc}{
short=EGC,
long= equal gain combining,
}
\DeclareAcronym{dft}{
short=DFT,
long= discrete Fourier transform,
}
\DeclareAcronym{kld}{
short=KLD,
long= Kullback–Leibler divergence,
}

%% file: figures/MCRB_vs_P.tex
\begin{tikzpicture}

\begin{axis}[%
width=2.75in,
height=1.4in,
at={(0in,0in)},
scale only axis,
xmin=-10,
xmax=40,
xlabel style={font=\color{white!15!black}},
xlabel={Transmitted power $P$ [dBm]},
ymode=log,
ymin=3e-03,
ymax=10,
ylabel style={font=\color{white!15!black}},
ylabel={RMSE [$\unit[]{m}$]},
yminorticks=true,
axis background/.style={fill=white},
xmajorgrids,
ymajorgrids,
yminorgrids,
legend style={at={(0,1)}, font=\scriptsize, anchor=north west, legend cell align=left, align=left, draw=white!15!black}
]

\addplot [color=blue, mark=x, mark size=3.3pt,mark options={solid, blue},line width=1pt]
  table[row sep=crcr]{%
  -10	0.889128502738626\\
  -5	0.482304265897224\\
  0	0.267935878248287\\
  5	0.151211587665345\\
  10	0.087441309573895\\
  15	0.0534627937329172\\
  20	0.0366025775267849\\
  25	0.0292287293399754\\
  30	0.0264058137774717\\
  35	0.0254083832056156\\
  40	0.0250512940124719\\
};
\addlegendentry{ML (pos. mismatch)}

\addplot [color=blue, dashed, mark=square, mark size=3pt,mark options={solid, blue},line width=1pt]
  table[row sep=crcr]{%
  -10	0.846403217503069\\
  -5	0.476410094926132\\
  0	0.268690565089602\\
  5	0.152484192639236\\
  10	0.0881716134558079\\
  15	0.0536648331087028\\
  20	0.0364992428807203\\
  25	0.0290303775376409\\
  30	0.0262295798717661\\
  35	0.025279392182549\\
  40	0.0249713930966303\\
};
\addlegendentry{LB (pos. mismatch)}

\addplot [color=blue, dashed,line width=1pt]
  table[row sep=crcr]{%
  -10	0.0248275677523335\\
  -5	0.0248276537125904\\
  0	0.0248276537125632\\
  5	0.0248276537125525\\
  10	0.0248276587133737\\
  15	0.0248276587133554\\
  20	0.0248276590197878\\
  25	0.0248276590187157\\
  30	0.0248276590197902\\
  35	0.0248276590188475\\
  40	0.0248276590197731\\
};
\addlegendentry{Bias (pos. mismatch)}

\end{axis}


\begin{axis}[%
  width=2.75in,
  height=1.4in,
  at={(0in,0in)},
  scale only axis,
  xmin=-10,
  xmax=40,
  xtick={\empty},
  ymode=log,
  ymin=3e-03,
  ymax=10,
  ytick={\empty},
  legend style={at={(1,1)}, font=\scriptsize, anchor=north east, legend cell align=left, align=left, draw=white!15!black, fill opacity=1}
  ]
  
  \addplot [color=red, mark=+, mark size=3.6pt,mark options={solid, red},line width=1pt]
    table[row sep=crcr]{%
    -10	0.945701796948886\\
    -5	0.537223830011796\\
    0	0.335916380883954\\
    5	0.24229257520225\\
    10	0.20368877021528\\
    15	0.189382237923837\\
    20	0.184299099982199\\
    25	0.182484961873893\\
    30	0.181830709725284\\
    35	0.181542339667633\\
    40	0.181439992687649\\
  };
  \addlegendentry{ML (ori. mismatch)}
  
  \addplot [color=red, dashed, mark=o, mark size=3.6pt,mark options={solid, red},line width=1pt]
    table[row sep=crcr]{%
    -10	0.886436425248452\\
    -5	0.520509349217151\\
    0	0.328821123219376\\
    5	0.237990224365415\\
    10	0.200894702764253\\
    15	0.18764403315994\\
    20	0.183254513314695\\
    25	0.181844375287026\\
    30	0.181396169117403\\
    35	0.181254203223822\\
    40	0.181209286519745\\
  };
  \addlegendentry{LB (ori. mismatch)}
  
  \addplot [color=red, dashed,line width=1pt]
    table[row sep=crcr]{%
    -10	0.181188271622172\\
    -5	0.181188509879586\\
    0	0.18118850982267\\
    5	0.18118850982265\\
    10	0.18118850984642\\
    15	0.18118850984635\\
    20	0.18118850984642\\
    25	0.181188509888167\\
    30	0.181188509888129\\
    35	0.18118850988818\\
    40	0.181188509888172\\
  };
  \addlegendentry{Bias (ori. mismatch)}
  
  \addplot [color=black,line width=1pt]
    table[row sep=crcr]{%
    -10	0.841193299768545\\
    -5	0.473037754932775\\
    0	0.266008677973854\\
    5	0.149587672483945\\
    10	0.084119329976739\\
    15	0.0473037754932892\\
    20	0.026600867797375\\
    25	0.0149587672483941\\
    30	0.00841193299768473\\
    35	0.00473037754933004\\
    40	0.00266008677974184\\
  };
  \addlegendentry{Mismatch-free PEB}
  
  \end{axis}

\end{tikzpicture}%

%% file: figures/MCRB_vs_sigma2.tex
%
%
\begin{tikzpicture}
  \begin{axis}[%
    width=2.75in,
    height=0.6in,
    at={(0in,0in)},
  scale only axis,
  xmin=0.09,
  xmax=0.61,
  xlabel style={font=\color{white!15!black}},
  xlabel={$\unit[\sigma_\mathrm{o}]{[{deg}]}$},
  ymode=log,
  ymin=0.05,
  ymax=1,
  yminorticks=true,
  ylabel style={font=\color{white!15!black}},
  ylabel={RMSE [$\unit[]{m}$]},
  axis background/.style={fill=white},
  xmajorgrids,
  ymajorgrids,
  yminorgrids,
  legend style={at={(0,1)}, font=\scriptsize, anchor=north west,legend cell align=left, align=left, draw=white!15!black}
  ]
  
  \addplot [color=blue, line width=1.0pt, mark=o, mark options={solid, blue}]
   plot [error bars/.cd, y dir=both, y explicit, error bar style={line width=1.0pt}, error mark options={line width=1.0pt, mark size=4.0pt, rotate=90}]
   table[row sep=crcr, y error plus index=2, y error minus index=3]{%
   0.1	0.0929907680490119	0.0292649245536898	0.00888393944481572\\
   0.2	0.113974157352882	0.082421703522655	0.029827416445583\\
   0.3	0.140753854905995	0.140167486679762	0.0565177753158361\\
   0.4	0.170574123207005	0.199258077500105	0.0862022961009108\\
   0.5	0.201909075310515	0.25822668200503	0.117358121336454\\
   0.6	0.234634959092299	0.321310957919717	0.14986452459521\\
  };
  \addlegendentry{LB (ori. mismatch)}
  
  \addplot [color=red, dashed, line width=1.5pt]
    table[row sep=crcr]{%
    0.1	0.084119329976739\\
    0.2	0.084119329976739\\
    0.3	0.084119329976739\\
    0.4	0.084119329976739\\
    0.5	0.084119329976739\\
    0.6	0.084119329976739\\
  };
  \addlegendentry{Mismatch-free PEB}
  
  \end{axis}

  \begin{axis}[%
    width=2.75in,
    height=0.6in,
    at={(0in,0in)},
  scale only axis,
  xmin=0.09,
  xmax=0.61,
  ymode=log,
  ymin=0.05,
  ymax=1,
  yminorticks=false,
  xtick={\empty},
  ytick={\empty},
  legend style={at={(0,1)}, font=\scriptsize, anchor=north west,legend cell align=left, align=left, draw=white!15!black}
  ]
  
  \addplot [color=blue, line width=1.0pt, mark=o, mark options={solid, blue}, forget plot]
   plot [error bars/.cd, y dir=both, y explicit, error bar style={line width=1.0pt}, error mark options={line width=1.0pt, mark size=4.0pt, rotate=90}]
   table[row sep=crcr, y error plus index=2, y error minus index=3]{%
   0.1	0.0929907680490119	0.0292649245536898	0.00888393944481572\\
   0.2	0.113974157352882	0.082421703522655	0.029827416445583\\
   0.3	0.140753854905995	0.140167486679762	0.0565177753158361\\
   0.4	0.170574123207005	0.199258077500105	0.0862022961009108\\
   0.5	0.201909075310515	0.25822668200503	0.117358121336454\\
   0.6	0.234634959092299	0.321310957919717	0.14986452459521\\
  };
  
  \addplot [color=red, dashed, line width=1.5pt, forget plot]
    table[row sep=crcr]{%
    0.1	0.084119329976739\\
    0.2	0.084119329976739\\
    0.3	0.084119329976739\\
    0.4	0.084119329976739\\
    0.5	0.084119329976739\\
    0.6	0.084119329976739\\
  };
  
  \end{axis}


\begin{axis}[%
  width=2.75in,
  height=0.6in,
  at={(0in,1.1in)},
scale only axis,
xlabel style={font=\color{white!15!black}},
xlabel={$\unit[\sigma_\mathrm{p}]{[m]}$},
xmin=0.009,
xmax=0.061,
ymode=log,
ymin=0.05,
ymax=1,
yminorticks=true,
ylabel style={font=\color{white!15!black}},
ylabel={RMSE [$\unit[]{m}$]},
xtick={0.01,0.02,0.03,0.04,0.05,0.06},
xmajorgrids,
ymajorgrids,
yminorgrids,
legend style={at={(0,1)}, font=\scriptsize, anchor=north west,legend cell align=left, align=left, draw=white!15!black}
]

\addplot [color=blue, line width=1.0pt, mark=o, mark options={solid, blue}]
 plot [error bars/.cd, y dir=both, y explicit, error bar style={line width=1.0pt}, error mark options={line width=1.0pt, mark size=4.0pt, rotate=90}]
 table[row sep=crcr, y error plus index=2, y error minus index=3]{%
 0.01	0.0932726753428712	0.0445579049873599	0.00893012007313974\\
 0.02	0.114329550360036	0.120476966430145	0.0293398761819626\\
 0.03	0.141018914976076	0.200869843716523	0.0549677946469987\\
 0.04	0.1707647646209	0.282896784223409	0.0832529491994101\\
 0.05	0.202277460076893	0.366531236581574	0.112925281843744\\
 0.06	0.234928442429327	0.4519889736101	0.143379130353293\\
};
\addlegendentry{LB (pos. mismatch)}

\addplot [color=red, dashed, line width=1.5pt]
  table[row sep=crcr]{%
  0.01	0.084119329976739\\
  0.02	0.084119329976739\\
  0.03	0.084119329976739\\
  0.04	0.084119329976739\\
  0.05	0.084119329976739\\
  0.06	0.084119329976739\\
};
\addlegendentry{Mismatch-free PEB}

\end{axis}

\begin{axis}[%
  width=2.75in,
  height=0.6in,
  at={(0in,1.1in)},
scale only axis,
xmin=0.009,
xmax=0.061,
ymode=log,
ymin=0.05,
ymax=1,
yminorticks=false,
xtick={\empty},
ytick={\empty},
xticklabels = {\empty},
yticklabels = {\empty},
legend style={at={(0,1)}, font=\scriptsize, anchor=north west,legend cell align=left, align=left, draw=white!15!black}
]

\addplot [color=blue, line width=1.0pt, mark=o, mark options={solid, blue}, forget plot]
 plot [error bars/.cd, y dir=both, y explicit, error bar style={line width=1.0pt}, error mark options={line width=1.0pt, mark size=4.0pt, rotate=90}]
 table[row sep=crcr, y error plus index=2, y error minus index=3]{%
 0.01	0.0932726753428712	0.0445579049873599	0.00893012007313974\\
 0.02	0.114329550360036	0.120476966430145	0.0293398761819626\\
 0.03	0.141018914976076	0.200869843716523	0.0549677946469987\\
 0.04	0.1707647646209	0.282896784223409	0.0832529491994101\\
 0.05	0.202277460076893	0.366531236581574	0.112925281843744\\
 0.06	0.234928442429327	0.4519889736101	0.143379130353293\\
};

\addplot [color=red, dashed, line width=1.5pt, forget plot]
  table[row sep=crcr]{%
  0.01	0.084119329976739\\
  0.02	0.084119329976739\\
  0.03	0.084119329976739\\
  0.04	0.084119329976739\\
  0.05	0.084119329976739\\
  0.06	0.084119329976739\\
};

\end{axis}

\end{tikzpicture}%